\documentclass[twocolumn,showpacs,aps,amssymb,floatfix,prd,amsmath,preprintnumbers]{revtex4}
\usepackage{epstopdf}
\usepackage{capt-of}
\usepackage{graphicx}  % Include figure files
\usepackage{dcolumn}
\usepackage{hyperref}   % Align table columns on decimal point
\usepackage{bm}% bold math
\begin{document}
\input epsf.tex
%%%%%%%%%%%%
%%%%%%%%%%%
\title{Non-exotic matter wormholes in a trace of the energy-momentum tensor squared gravity}

\author{
    P.H.R.S. Moraes$^{*}$, P.K. Sahoo$^{\dagger}$}
    \affiliation{$^{*}$ITA - Instituto Tecnol\'ogico de Aeron\'autica - Departamento de F\'isica, 12228-900, S\~ao Jos\'e dos Campos, S\~ao Paulo, Brazil \footnote{Email: moraes.phrs@gmail.com}}
    \affiliation{$^{\dagger}$Department of Mathematics, Birla Institute of Technology and Science-Pilani, \\ Hyderabad Campus, Hyderabad-500078, India \footnote{Email: pksahoo@hyderabad.bits-pilani.ac.in}}

\begin{abstract}

Wormholes are tunnels connecting two different points in space-time. In Einstein's General Relativity theory, wormholes are expected to be filled by exotic matter, i.e., matter that does not satisfy the energy conditions and may have negative density. We propose, in this paper, the achievement of wormhole solutions with no need for exotic matter. In order to achieve so, we consider a gravity theory that starts from linear and quadratic terms on the trace of the energy-momentum tensor in the gravitational action. We show that by following this formalism, it is possible, indeed, to obtain non-exotic matter wormhole solutions.

\end{abstract}


%%%%%


\maketitle

%%%%%%%%%%%%%%%%%%%%%%%%%%%%%%%%%%%%%%%%%%

\section{Introduction}

Extended gravity theories (EGTs) have been constantly proposed and worked out with the main purpose of solving or at least evading some observational issues in cosmology \cite{padmanabhan/2003,capozziello/2008,bull/2016}. From the Einstein's field equations of General Relativity (GR), which in natural units, that shall be assumed throughout the present article, read $G_{\mu\nu}=8\pi T_{\mu\nu}$, with $G_{\mu\nu}$ and $T_{\mu\nu}$ being the Einstein and energy-momentum tensors, respectively, we can note that in order to extend gravity, one has to correct $G_{\mu\nu}$ (geometrical corrections) or $T_{\mu\nu}$ (material corrections).

When generalizing the geometrical side of Einstein's field equations, the most popular models found are those derived from the $f(R)$ theories \cite{sotiriou/2010,de_felice/2010}, with $f(R)$ indicating generic functions of the Ricci scalar $R$. One can also find models that consider the Gauss-Bonnet term, $G$, or functions of it, $f(G)$, in the gravitational part of the action, yielding the Gauss-Bonnet models \cite{nojiri/2005,ivashchuk/2016}. Naturally, it is also possible to further generalize the geometrical aspect of a gravitational theory by considering generic terms on both $R$ and $G$, as made in the $f(R,G)$ theories \cite{bamba/2010,de_laurentis/2015}.

On the other side, the generalization of the material sector of the Einstein's field equations may come from quintessence models \cite{chimento/2003,ms/2014}, in which the dynamics of the universe is considered to be governed by one or more arbitrary scalar fields.

Recently, attention has also been given to models able to generalize both geometrical and material sectors of a gravitational theory at the same time, for which we quote the $f(R,L_m)$ and $f(R,T)$ theories, that have been respectively proposed in \cite{harko/2010,harko/2011}, in which $L_m$ is the matter lagrangian and $T$ is the trace of the energy-momentum tensor.

In the present article we will focus our attention on the $f(R,T)$ theory of gravity. This theory has shown to provide a good alternative to the cosmological issues quoted above, as it can be seen, for instance, in References \cite{ms/2017,shabani/2017,zaregonbadi/2016,baffou/2015,sun/2016}. Other interesting references that show the wide applicability of $f(R,T)$ gravity are \cite{das/2017,das/2016,mam/2016,amam/2016}.

Another utility of EGTs may come from the study of wormholes (WHs). WHs are theoretically predicted as a particular case of GR's Schwarzschild solution that does not contain horizons and can connect two different points in space-time through a tunnel filled by exotic matter, that violates the energy conditions \cite{morris/1988,morris/1988b}. 

WHs were firstly introduced as a tool for teaching GR. Over time, the interest in such objects has grown and since the late 90's of the last century it is common to see in the literature attempts to search for observational signatures of WHs \cite{zhou/2016,harko/2008,safonova/2002}.

One may wonder what would be the advantages of modelling WHs in EGTs. As quoted above, when assuming GR as the background gravitational theory, the physical content of a traversable WH is somehow exotic. In other words, the matter threading these objects violates the energy conditions. 

The energy conditions are based on the Raychaudhuri equation and are commonly used to stablish and study singularities in the space-time \cite{visser/1995}. Matter inside GR WHs would violate, for instance, the null energy condition (NEC) $T_{\mu\nu}u^\mu u^\nu\geq0$, with $u_{\mu}$ being some null vector.

It is natural to think of EGTs as having field equations as $G_{\mu\nu}=8\pi T_{\mu\nu}^{eff}$, with $T_{\mu\nu}^{eff}$ being defined as the bare matter energy-momentum tensor $T_{\mu\nu}$ plus correction terms. Following this approach, one can see that the correction terms in a given EGT may allow the NEC to be satisfied as $T_{\mu\nu}u^\mu u^\nu\geq0$. As a matter of fact,  non-exotic matter WHs have already been modelled in EGTs, as one can see in Refs. \cite{harko/2013,mazhamirousavi/2016,hohmann/2014}.

Our main goal in the present article is to construct non-exotic matter WHs, however, departing from the references just cited above, from EGTs that present corrections to the energy-momentum tensor, instead of to the Einstein tensor. A further theory with this aspect was recently proposed in \cite{roshan/2016} and allows the existence of a term proportional to $T_{\mu\nu}T^{\mu\nu}$ in the Einstein-Hilbert action. It has been shown that this theory avoids the existence of an early-time singularity and possesses a true sequences of cosmological eras.

Here, particularly, we will insert linear and quadratic terms in the trace of the energy-momentum tensor in the Einstein-Hilbert gravitational action, so that our background theory will be the $f(R,T)$ gravity \cite{harko/2011}, with $f(R,T)=R+h(T)$ and $h(T)=\alpha T+\beta T^{2}$, with $\alpha$ and $\beta$ constants. 

\section{Trace of the energy-momentum tensor squared gravity}\label{sec:sg} 

Let us take as our starting point, the following action 
\begin{equation}\label{sg1}
\mathcal{S}=\frac{1}{16\pi}\int d^{4}x\sqrt{-g}[R+h(T)]+\int d^{4}x\sqrt{-g}\mathcal{L}_m,
\end{equation}
with $g$ being the determinant of the metric $g_{\mu\nu}$, $h(T)$ a function of $T$ only and $\mathcal{L}_m$ the matter lagrangian. The motivation for inserting $h(T)$ in the Einstein-Hilbert action is the possible existence of imperfect fluids in the universe \cite{harko/2011}.

Following the steps of Ref. \cite{harko/2011}, the variational principle applied to (\ref{sg1}) yields the field equations 
\begin{equation}\label{sg2}
G_{\mu\nu}=8\pi T_{\mu\nu}+\frac{1}{2}h(T)g_{\mu\nu}+h^{\dagger}(T)(T_{\mu\nu}+\mathcal{P}g_{\mu\nu}),
\end{equation}
with $G_{\mu\nu}$ being the usual Einstein tensor, $T_{\mu\nu}=diag(\rho,-p_r,-p_t,-p_t)$, with $\rho$ being the matter-energy density, $p_r$ and $p_t$ the radial and transverse components of the pressure, $h^{\dagger}(T)\equiv dh(T)/dT$ and $\mathcal{P}=(p_r+2p_t)/3$ is the total pressure. Note that since the form we assumed for $T_{\mu\nu}$ is not invariant under a change of basis, it will be valid only in the basis that point in these radial and transverse privileged directions.

It is worth to remark that in the derivation process of Eq.(\ref{sg2}), we have also assumed $\mathcal{L}_m=-\mathcal{P}$ in Eq.(\ref{sg1}), as made in Ref. \cite{harko/2011}. It is known that this is not the unique choice for the matter lagrangian. On this regard, the theories proposed in \cite{harko/2010,harko/2011} predict an extra force to act perpendicularly to the $4-$velocity. The extra force implies that the movement of test particles in gravitational fields is non-geodesic. It has been shown in \cite{bertolami/2008} and later revisited in \cite{harko/2014} that when one assumes the matter lagrangian to be proportional to the total pressure, the extra force vanishes.

With the purpose of checking the possibility of modelling non-exotic matter WHs, we will take $h(T)=\alpha T+\beta T^{2}$.  It is interesting to recall that very recently, a theory powered by a squared energy-momentum tensor was proposed in \cite{roshan/2016}. These kind of theories contrast with most studies of higher-order gravity, since those focus on generalising the Einstein-Hilbert geometrical contribution to the lagrangian. It was shown that cosmological models derived from this idea may evade the Big-Bang singularity \cite{board/2017}. Also, the cosmic acceleration appears naturally in a matter-dominated universe in energy-momentum tensor powered gravity theories \cite{akarsu/2017}. Moreover, such a specific functional form for $h(T)$ was firstly introduced in a cosmological context in \cite{mrc/2016}. The results have described a universe that transits from a decelerated to an accelerated regime of expansion and agreed with observational data.

This assumption yields, for Eq.(\ref{sg2}), the following
\begin{equation}\label{sg3}
G_{\mu\nu}=8\pi T_{\mu\nu}^{eff},
\end{equation}
with 
\begin{widetext}
\begin{equation}\label{sg4}
T_{\mu\nu}^{eff}=T_{\mu\nu}+\frac{1}{8\pi}\left\{\alpha\left[T_{\mu\nu}+\frac{1}{2}(\rho-\mathcal{P})g_{\mu\nu}\right]+2\beta(\rho-3\mathcal{P})\left[T_{\mu\nu}+\frac{1}{4}(\rho+\mathcal{P})g_{\mu\nu}\right]\right\}.
\end{equation}
\end{widetext}

Note that we have written the field equations of the theory in terms of an effective energy-momentum tensor for the sake of the applicability of the energy conditions that will be made later.

Next, we will develop the above field equations for the WH metric.

\section{Wormholes in a trace of the energy-momentum tensor squared gravity}\label{sec:wh}

The WH metric, presented firstly in the seminal reference \cite{morris/1988}, reads

\begin{equation}\label{wh1}
ds^{2}=e^{2\Phi(r)}dt^{2}-\left[1-\frac{b(r)}{r}\right]^{-1}dr^{2}-r^{2}(d\theta^{2}+\sin^{2}\theta d\phi^{2}).
\end{equation}
In Eq.(\ref{wh1}), since we assume the absence of event horizons, the redshift function $\Phi(r)$ must be finite everywhere. 

By inserting metric (\ref{wh1}) in field equations (\ref{sg3})-(\ref{sg4}), we have

\begin{widetext}
\begin{equation}\label{wh3}
\frac{b'}{r^{2}}=8\pi\left\{\rho+\frac{1}{16\pi}\left[\alpha\left(3\rho-\frac{p_r+2p_t}{3}\right)+\beta(\rho-p_r-2p_t)\left(5\rho+\frac{p_r+2p_t}{3}\right)\right]\right\},
\end{equation}
\begin{equation}\label{wh4}
\frac{1}{r}\left[\frac{b}{r^2}+2\Phi'\left(\frac{b}{r}-1\right)\right]=8\pi\left\{-p_r+\frac{1}{16\pi}\left[\alpha\left(\rho-\frac{7p_r+2p_t}{3}\right)+\beta(\rho-p_r-2p_t)\left(\rho-\frac{11p_r-2p_t}{3}\right)\right]\right\},
\end{equation}
\begin{eqnarray}\label{wh5}
\frac{1}{2r}\left[\frac{1}{r}\left(\Phi'b+b'-\frac{b}{r}\right)+2(\Phi''+\Phi'^2)b-\Phi'(2-b')\right]-(\Phi''+\Phi'^2)\nonumber\\ =8\pi\left\{-p_t+\frac{1}{16\pi}\left[\alpha\left(\rho-\frac{p_r+8p_t}{3}\right)+\beta(\rho-p_r-2p_t)\left(\rho+\frac{p_r-10p_t}{3}\right)\right]\right\},
\end{eqnarray}
\end{widetext}
with primes denoting derivatives with respect to the coordinate $r$.

By specifying the redshift and shape functions in Eqs.(\ref{wh3})-(\ref{wh5}), one deduces the matter content of the WH. We will take the redshift function to be constant, as $\Phi'(r)=0$. This assumption has been deeply invoked in the literature, as one can check, among many others, Ref. \cite{zubair/2016}. Effectively, a constant redshift function can be absorbed in the rescaled time coordinate.

In order to obtain solutions for the matter content of WHs, it is also useful to invoke an equation of state (EoS), i.e., a relation between matter-energy density and pressure. Here we will use the well-known barotropic form for the EoS, which reads \cite{mustapha/2015}

\begin{eqnarray}
p_r=m\rho, \label{wh6} \\
p_t=n\rho, \label{wh6.1}
\end{eqnarray}
where the constants $m$ and $n$ are the EoS parameters.

Substituting (\ref{wh6})-(\ref{wh6.1}) in (\ref{wh3})-(\ref{wh5}), we have

\begin{widetext}
\begin{equation}\label{wh7}
\frac{b'}{r^{2}}=-\frac{1}{6} \{\alpha  [(m+2 n-9)]+\beta  (m+2 n-1) (m+2 n+15)\rho-48 \pi \}\rho,
\end{equation}
\begin{equation}\label{wh8}
\frac{b}{r^3}=\frac{1}{6} [3 \alpha -7 \alpha  m-2 \alpha  n+\beta  (m+2 n-1) (11 m-2 n-3)\rho-48 \pi  m]\rho,
\end{equation}
\begin{equation}\label{wh9}
\frac{1}{r^2}\left(b'-\frac{b}{r}\right)=-\frac{1}{3} \left\{\alpha  (m-3)+\beta  \left(m^2+2 m-3\right) \rho +8 n [\alpha -\beta  (m-2) \rho +6 \pi ]-20 \beta  n^2 \rho \right\}\rho.
\end{equation}
\end{widetext}

Next, we will obtain solutions for the matter content of WHs in a trace of the energy-momentum tensor squared gravity. In order to have non-constant matter-energy density solutions we will consider the shape function as 

\begin{equation}\label{shape}
b(r)=r\left(\frac{r_0}{r}\right)^{i+1}
\end{equation}
with constant $i$, such that its forms for different values of $i$ can be seen in Table 1. Our values for $i$, named $-0.5,0.5,1$, were used in References \cite{lobo/2009,Pavlovic/2015}.
 
\begin{table}[h]
\begin{tabular}{|c|c|c|c|}
  \hline
  $i$ & 1 & 0.5 & -0.5 \\\hline
  $b(r)$ & $r_0^2/r$ & $r_0\sqrt{r_0/r}$ & $\sqrt{r_0r}$ \\
   \hline
\end{tabular}
\caption{Shape function (\ref{shape}) for different values of $i$.}
\end{table}

\begin{figure}[h]
\centering
\includegraphics[width=75mm]{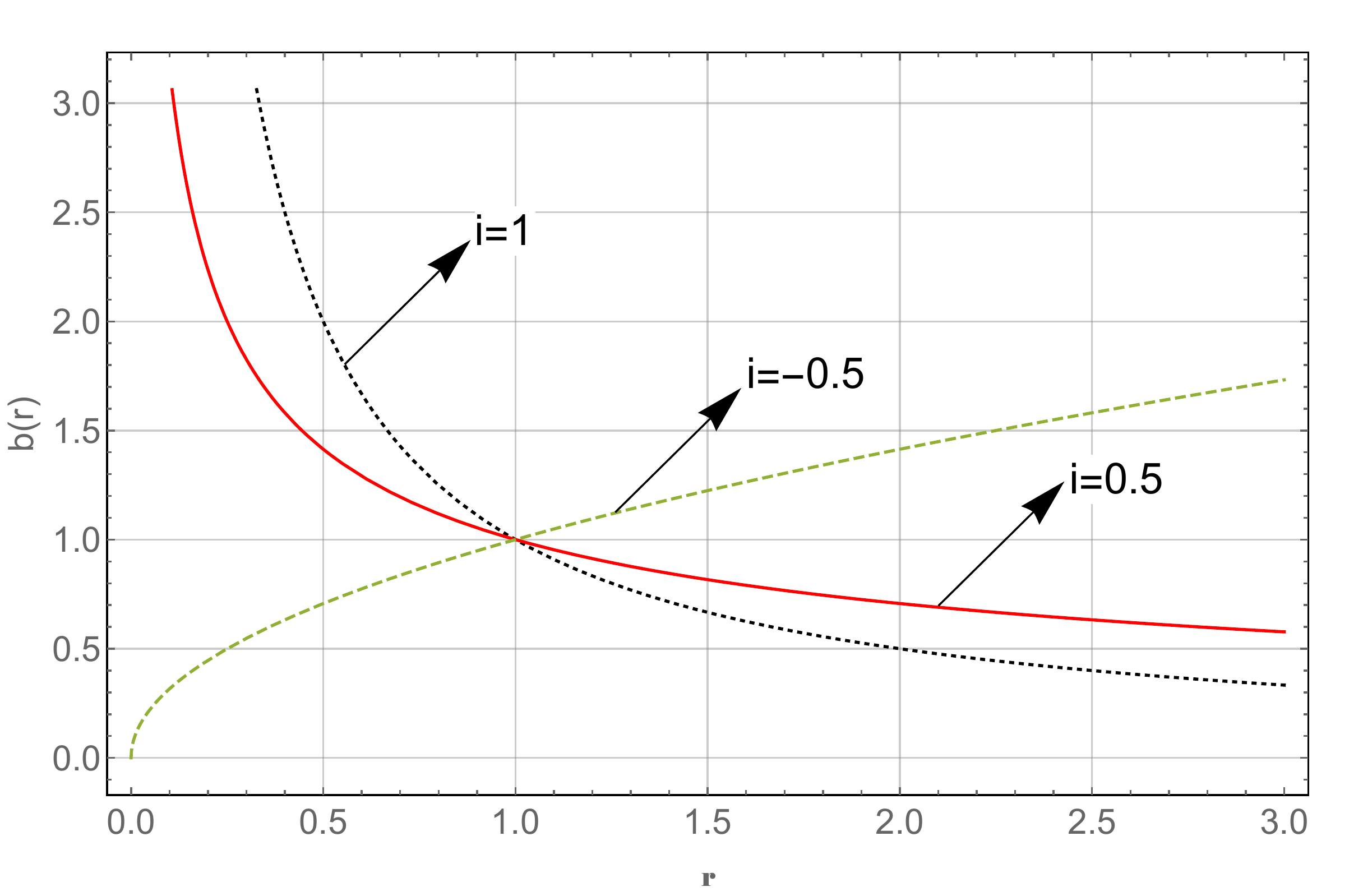}
\caption{Behaviour of the shape function $b(r)$ with $r_0=1$.}\label{figI}
\end{figure}

Figure \ref{figI} represents the behaviour of the shape function $b(r)$ versus $r$. For $i=-0.5$ the shape function  increases and for the other two cases, it decreases with radius $r$.

\begin{figure}[h]
\centering
\includegraphics[width=75mm]{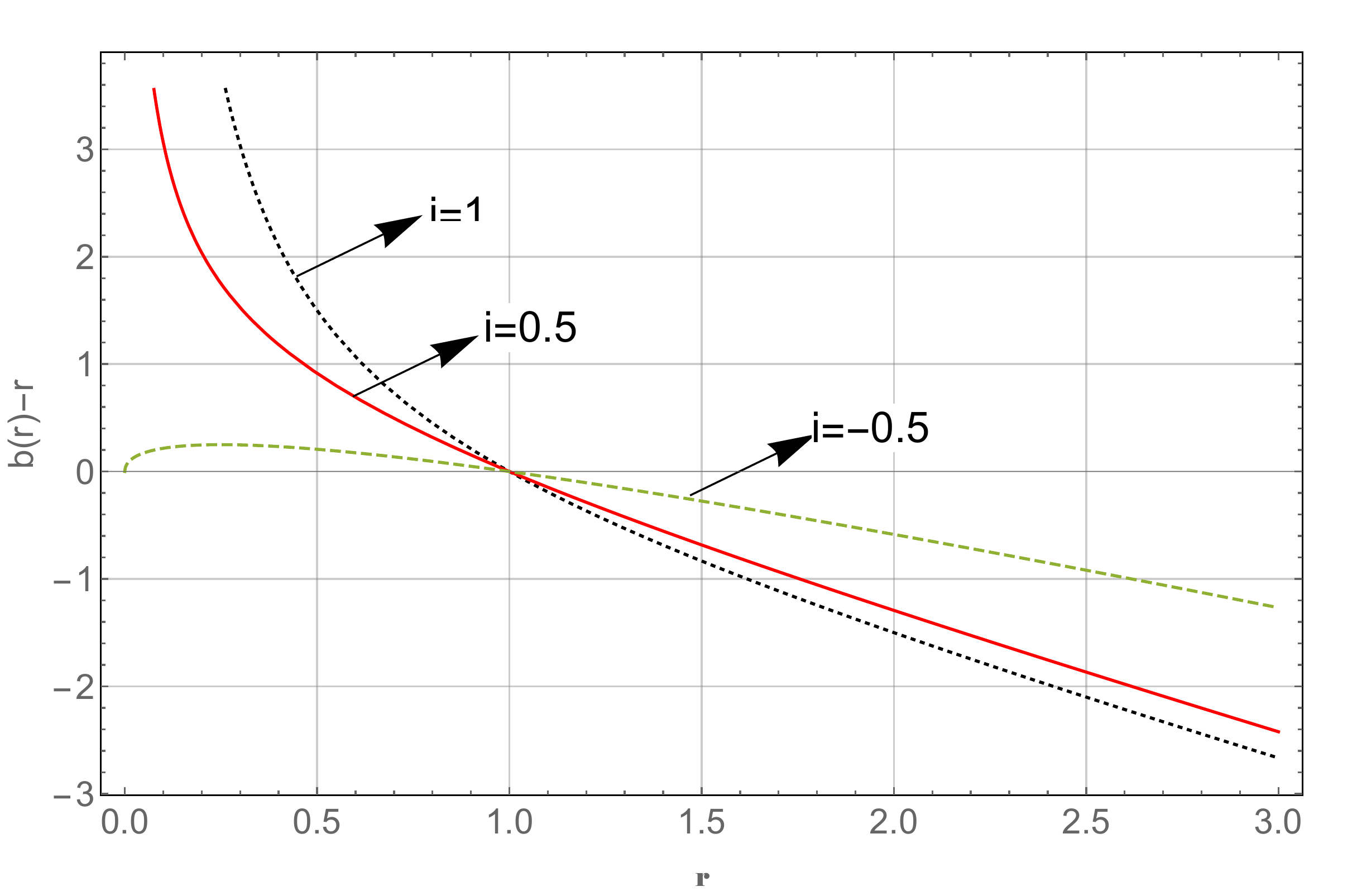}
\caption{Profile of $b(r)-r$ versus $r$ with $r_0=1$ for different values of $i$.}\label{figII}
\end{figure}

The throat of the WH occurs where the graph of $b(r)-r$ cuts the $r$ axis and for all the three cases the throat occurs at $r=1$ as it can be seen in Fig.\ref{figII}. The graph clearly indicates that $b(r)-r <0$ for $r>r_0$, satisfying the necessary metric condition $1-b(r)/r >0$ \cite{morris/1988}. Fig.\ref{figII} also agrees with the requirement $b(r_0)=r_0$ \cite{morris/1988} with $r_0=1$.
 
\begin{figure}[h]
\centering
\includegraphics[width=75mm]{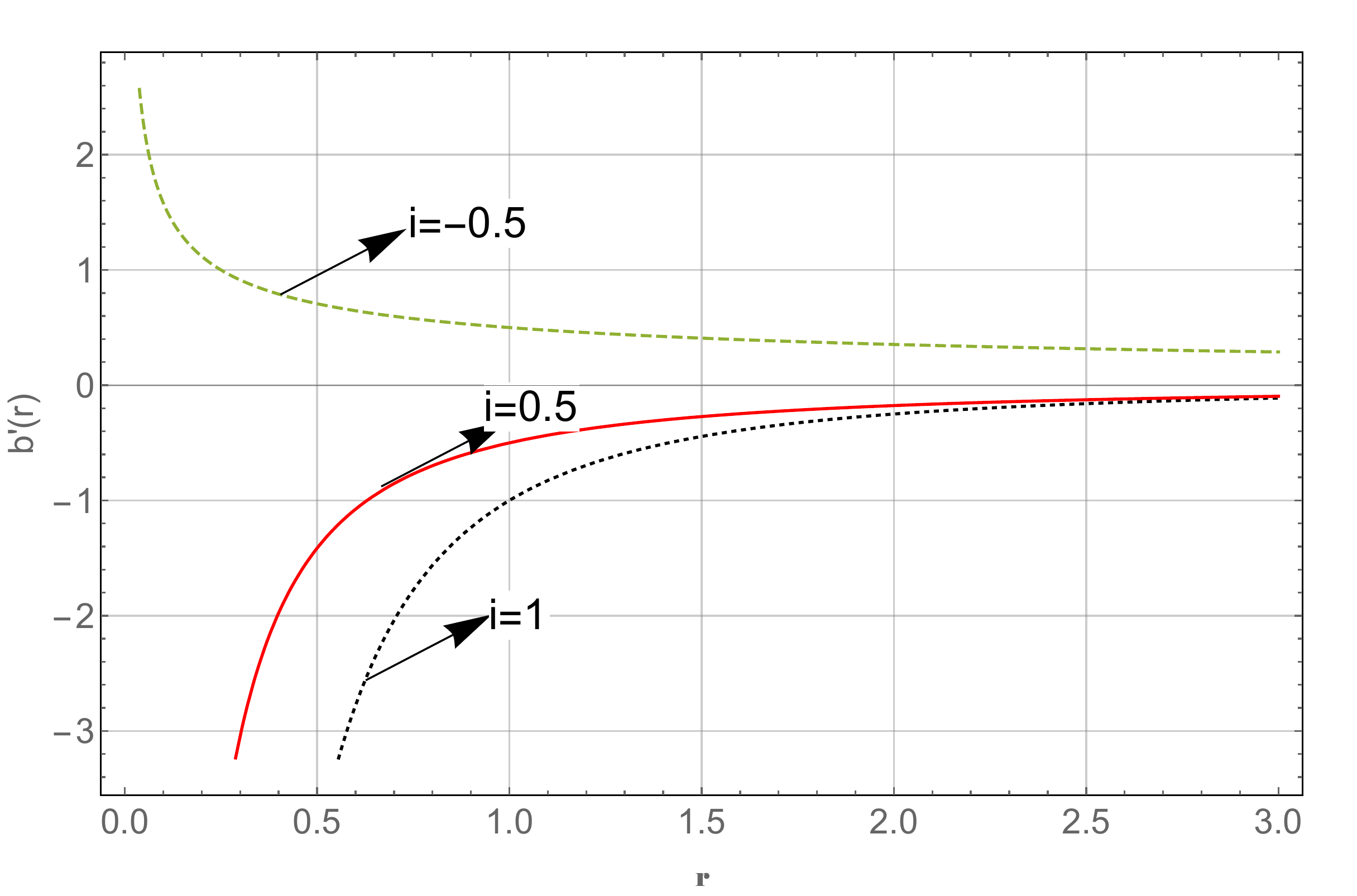}
\caption{Behaviour of the radial derivative of the shape function with $r_0=1$ for different values of $i$.}\label{figIII}
\end{figure}

Figure \ref{figIII} depicts the radial derivative of the shape function, which indicates $b'(r_0)< 1$ in all cases, as it is necessary \cite{morris/1988}. Hence, the shape functions satisfy the required structure conditions for WHs.

In the next section we will obtain the WH material content solutions for a particular value of $i$ and apply the energy conditions to them.

\section{The case $i=0.5$ - Matter content solutions and energy conditions}\label{sec:EC}

Motivated by Reference \cite{Pavlovic/2015}, in which WHs in $f(R)$ models were investigated, we consider $i=0.5$ in Eq.(\ref{shape}), yielding
  
\begin{equation}\label{ex2}
b=r_0\sqrt{\frac{r_0}{r}}.
\end{equation}

Substituting (\ref{ex2}) in (\ref{wh8}) we have

\begin{widetext}
\begin{equation}\label{wh13}
\rho=\frac{\sqrt{24 \beta  \left(\frac{r_0}{r}\right)^{3/2} (m+2 n-1) (11 m-2 n-3)+r^2 [\alpha  (7 m+2 n-3)+48 \pi  m]^2}}{2 \beta  r (11 m-2 n-3) (m+2 n-1)}+\frac{\alpha  (7 m+2 n-3)+48 \pi  m}{2 \beta  (11 m-2 n-3) (m+2 n-1)}.
\end{equation}
\end{widetext}

\begin{figure}[h!]
\centering
\includegraphics[width=75mm]{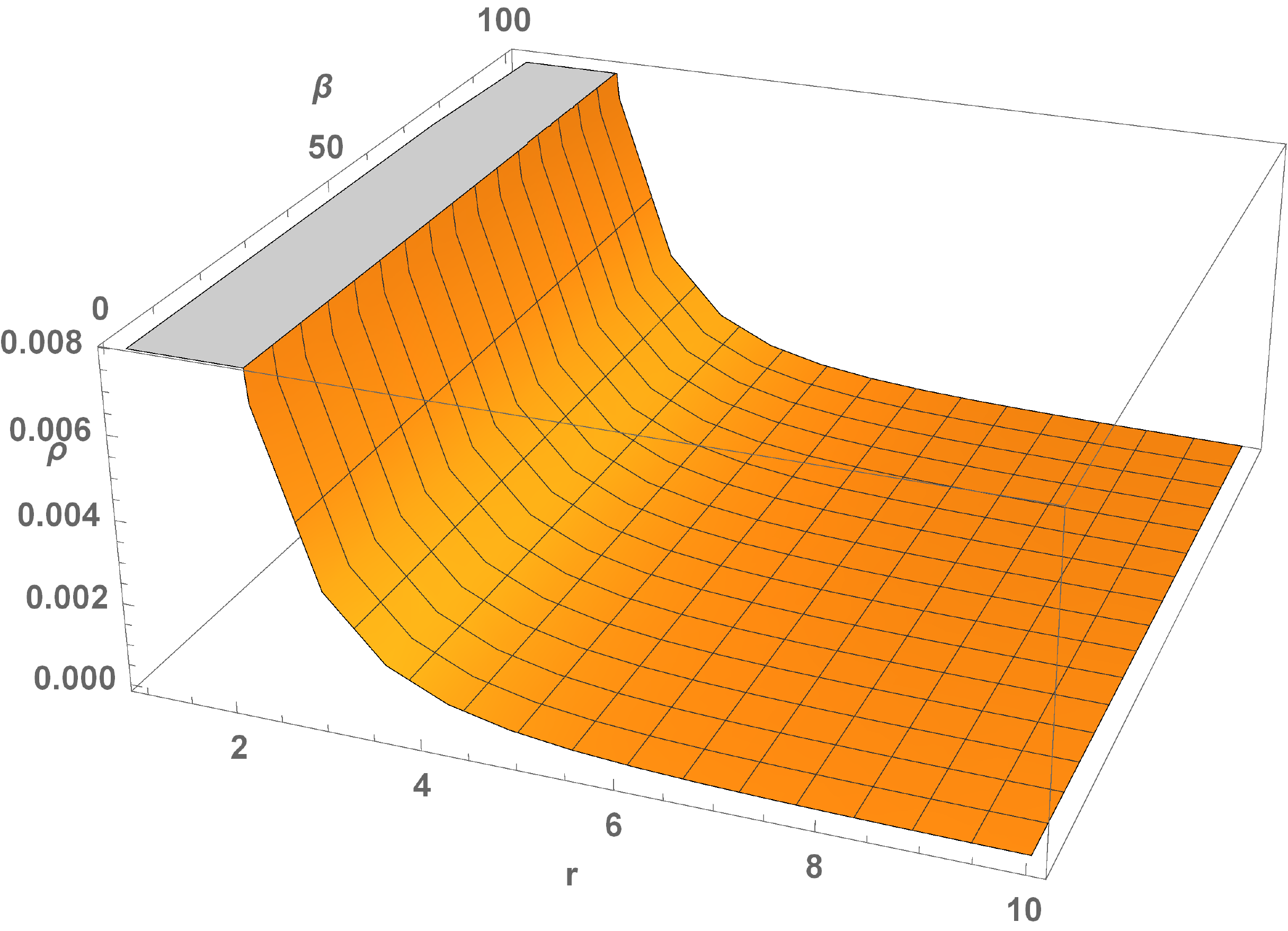}
\caption{Behaviour of $\rho$ with $r_0=1$, $m=0.5$, $n=1.5$ and $\alpha=-35$.}\label{fig1}
\end{figure}

\begin{figure}[h!]
\centering
\includegraphics[width=75mm]{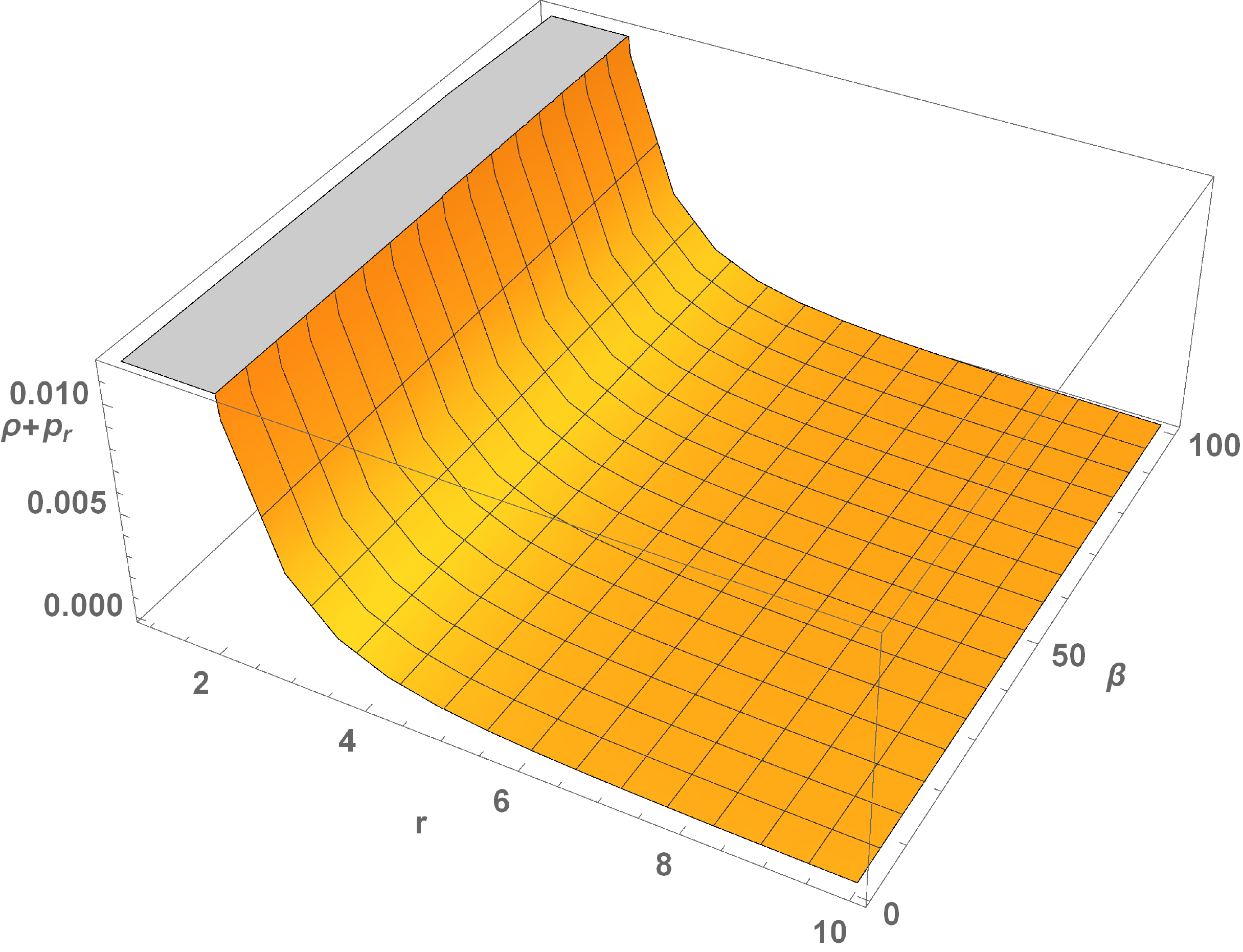}
\caption{Behaviour of NEC, $\rho+p_r$, with $r_0=1$, $m=0.5$, $n=1.5$ and $\alpha=-35$.}\label{fig3}
\end{figure}

\begin{figure}[h!]
\centering
\includegraphics[width=75mm]{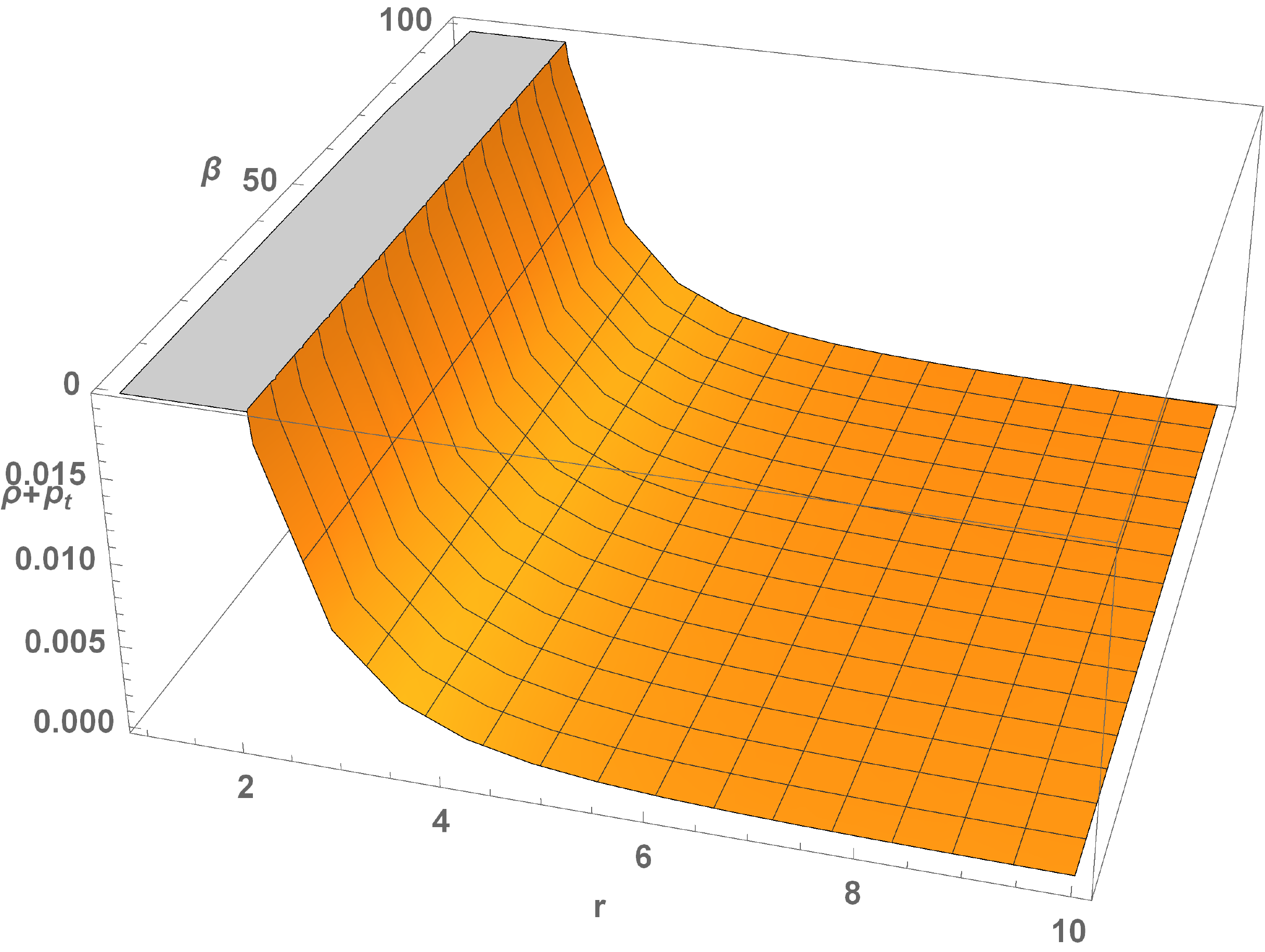}
\caption{Behaviour of NEC, $\rho+p_t$, with $r_0=1$, $m=0.5$, $n=1.5$ and $\alpha=-35$.}\label{fig5}
\end{figure}

\begin{figure}[h!]
\centering
\includegraphics[width=75mm]{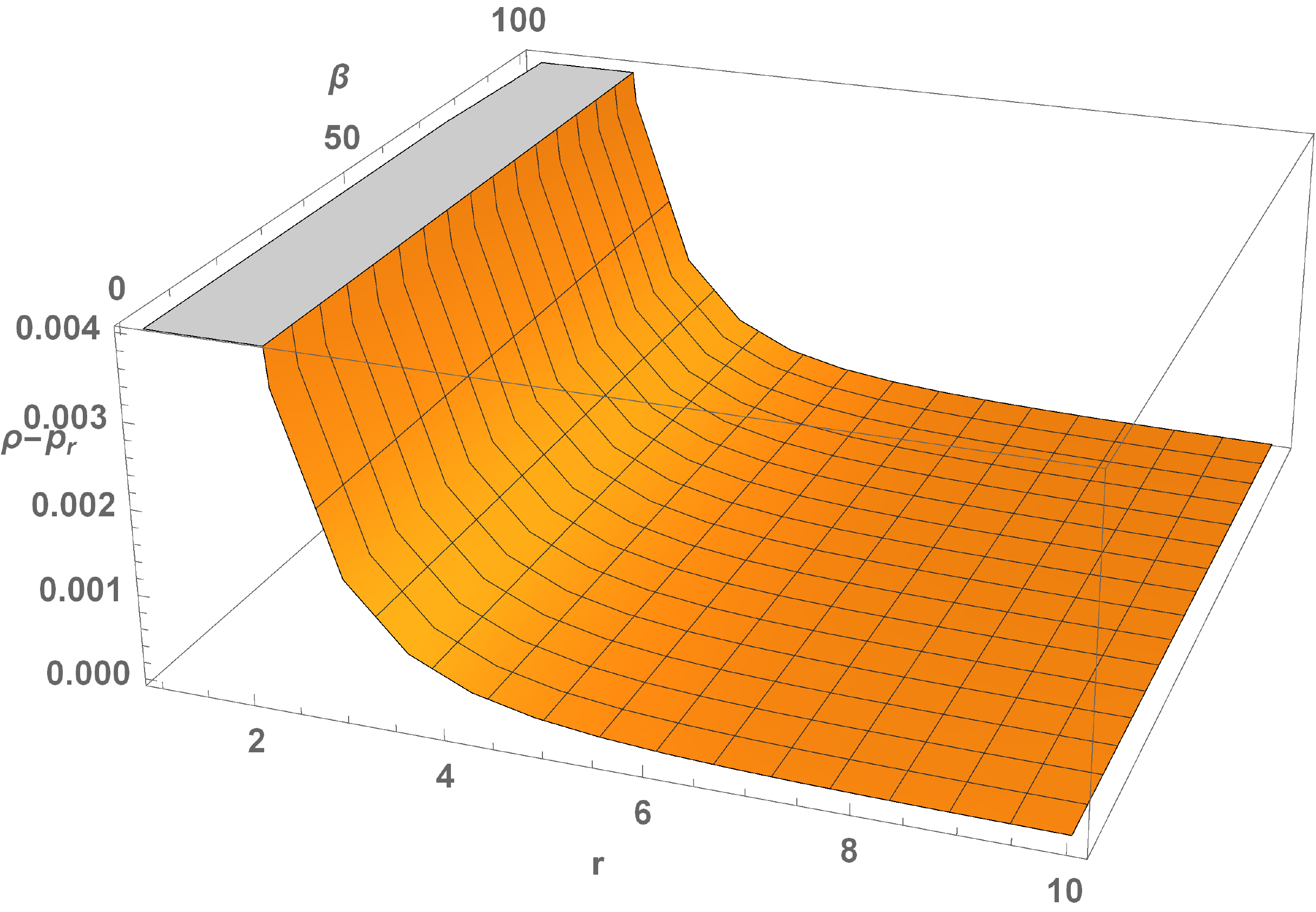}
\caption{Behaviour of DEC, $\rho-p_r$, with $r_0=1$, $m=0.5$, $n=1.5$ and $\alpha=-35$.}\label{fig7}
\end{figure}

\begin{figure}[h!]
\centering
\includegraphics[width=75mm]{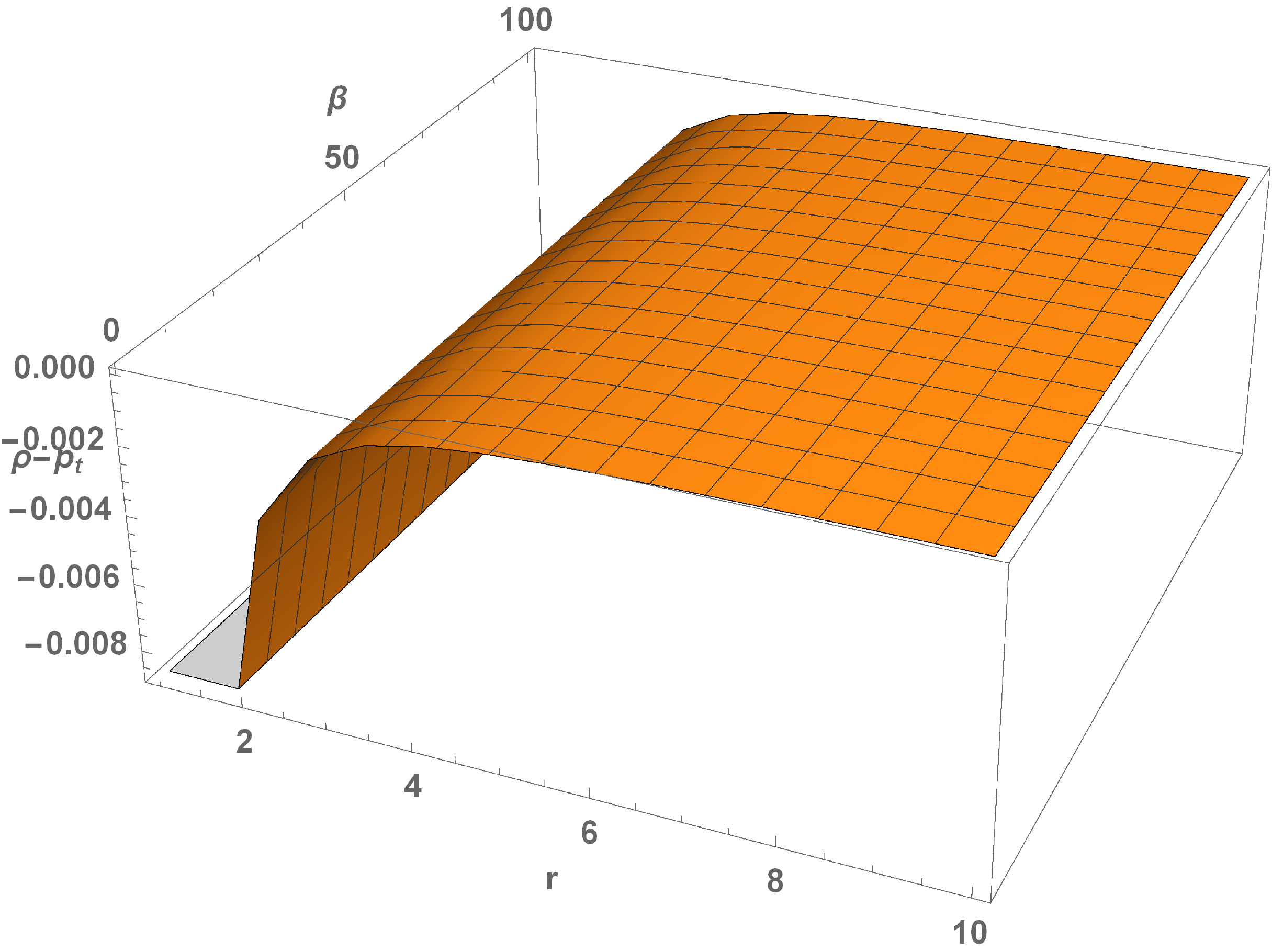}
\caption{Behaviour of DEC, $\rho-p_t$, with $r_0=1$, $m=0.5$, $n=1.5$ and $\alpha=-35$.}\label{fig9}
\end{figure}

\begin{figure}[h!]
\centering
\includegraphics[width=75mm]{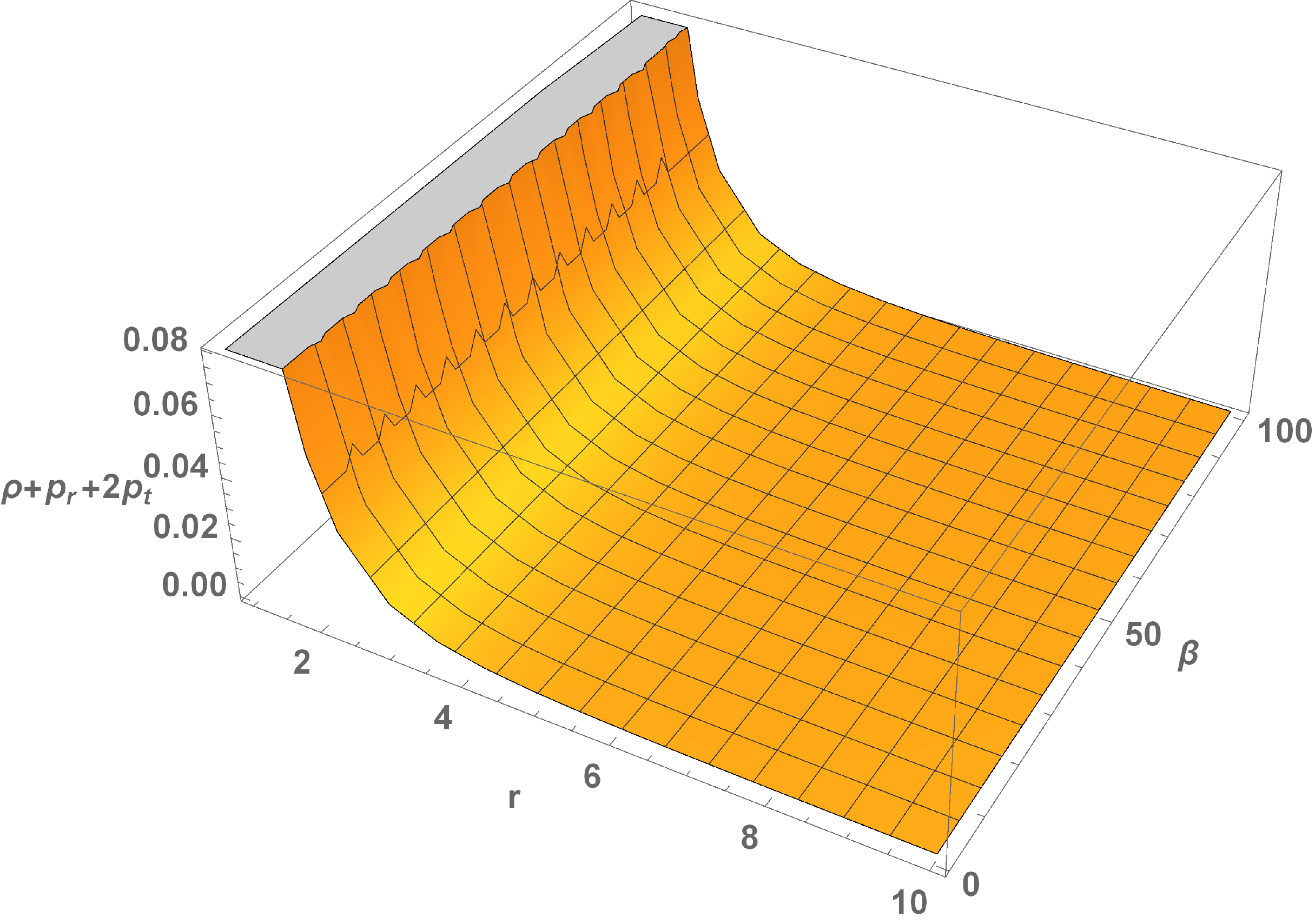}
\caption{Behaviour of SEC, $\rho+p_r+2p_t$, with $r_0=1$, $m=0.5$, $n=1.5$ and $\alpha=-35$.}\label{fig11}
\end{figure}

In Figures \ref{fig1} below we depict the behaviour of solution (\ref{wh13}). Since we are mainly concerned with the effects generated by the quadratic material terms of the theory, we will fix the value of $\alpha$ and plot $\rho$ in a wide range for $\beta$. Moreover, $m$ and $n$ are set equal to $0.5$ and $1.5$, respectively. The same approach will be assumed in Figs.\ref{fig3}-\ref{fig11}.

%\begin{figure}[h!]
%\centering
%\includegraphics[width=75mm]{11.pdf}
%\caption{Behaviour of $\rho$ with $r_0=1$, $m=0.5$, $n=1.5$ and $\beta=15$.}\label{fig2}
%\end{figure}

In possession of Eq.(\ref{wh13}), we are also able to plot the energy conditions of the WH matter content with respect to the radial coordinate $r$. Let us start with the NEC. It states that $\rho+p_r\geq0$ and $\rho+p_t\geq0$ \cite{visser/1995}. From Figures \ref{fig3} and \ref{fig5}, one can observe that the model validates NEC for positive $\beta$.

%\begin{figure}[h!]
%\centering
%\includegraphics[width=75mm]{22.pdf}
%\caption{Behaviour of NEC, $\rho+p_r$, with $r_0=1$, $m=0.5$, $n=1.5$ and $\beta=155$.}\label{fig4}
%\end{figure}

%\begin{figure}[h!]
%\centering
%\includegraphics[width=75mm]{33.pdf}
%\caption{Behaviour of NEC, $\rho+p_t$, with $r_0=1$, $m=0.5$, $n=1.5$ and $\beta=155$.}\label{fig6}
%\end{figure}

The weak energy condition (WEC) states that NEC should be satisfied along with $\rho\geq0$ \cite{visser/1995}. In this manner, from Figs.\ref{fig1}, \ref{fig3} and \ref{fig5}, we can see that WEC is satisfied.

From Figures \ref{fig7} and \ref{fig9} below, we can observe that the dominant energy condition (DEC), $\rho-p_r\geq0$ and $\rho-p_t\geq0$ \cite{visser/1995}, is valid for radial pressure and violated for transversal pressure, for $\beta>0$.

%\begin{figure}[h!]
%\centering
%\includegraphics[width=75mm]{44.pdf}
%\caption{Behaviour of DEC, $\rho-p_r$, with $r_0=1$, $m=0.5$, $n=1.5$ and $\beta=15$.}\label{fig8}
%\end{figure}

%\begin{figure}[h!]
%\centering
%\includegraphics[width=75mm]{55.pdf}
%\caption{Behaviour of DEC, $\rho-p_t$, with $r_0=1$, $m=0.5$, $n=1.5$ and $\beta=15$.}\label{fig10}
%\end{figure}

The strong energy condition (SEC) is the assertion that $\rho+p_r+2p_t\geq0$ \cite{visser/1995}. Figure \ref{fig11} shows that SEC is satisfied for positive $\beta$.

%\begin{figure}[h!]
%\centering
%\includegraphics[width=75mm]{66.pdf}
%\caption{Behaviour of SEC, $\rho+p_r+2p_t$, with $r_0=1$, $m=0.5$, $n=1.5$ and $\beta=15$.}\label{fig12}
%\end{figure}

\section{Conclusions}\label{sec:c}

Although EGTs with higher order curvature terms ($f(R)$, $f(G)$ and $f(R,G)$ theories) are most popular nowadays, gravity theories predicting corrections to the energy-momentum tensor have shown highly satisfactory results in Cosmology and Astrophysics. On this regard, besides the references quoted in Introduction, one may also check \cite{azevedo/2016}.

Corrections to the usual energy-momentum tensor of Einstein's field equations are mainly motivated by: i) the existence of imperfect fluids in the universe. In this sense, the extra terms in the effective energy-momentum tensor may work as terms of anisotropy or viscosity added to a perfect fluid; ii) quantum effects, such as particle creation. This possibility may be a clue that there is a connection between these EGTs and quantum theory of gravity. Gravitational induced particle production in  EGTs can be appreciated in \cite{harko/2015}.

Following the last paragraph, the modelling of WHs in this kind of EGTs is well motivated once the matter content of these objects is described by an anisotropic energy-momentum tensor. Even the creation of particles, described by the non-nullity of the covariant derivative of the energy-momentum tensor in these theories, was already analysed in WHs \cite{pan/2015,kim/1992}.

In the present article we have derived solutions to static Morris-Thorne WHs (\ref{wh1}) within an EGT that admits quadratic material terms in its field equations (\ref{sg3})-(\ref{sg4}). 

We have obtained WH solutions for three different cases of the shape function $b(r)$. It is important to stress here that all the functions $b(r)$ satisfy the necessary conditions. That is, the functions $b(r)$ reduce to $r_0$ at the WH throat and $b(r)/r<1$ for all $r>r_0$. One can also verify that the asymptotic behaviour $[b(r)/r]\rightarrow0$ as $r\rightarrow\infty$ is obeyed, as well as the flaring out condition $[b(r)-b'(r)r]/2b^{2}(r)>0$.

Among the examples found in the literature that present non-exotic matter WHs, one could quote \cite{hohmann/2014,maeda/2008}. Our main motivation in this paper was the possibility of obtaining WHs with non-exotic matter content, rather, in an EGT that allows material, instead of geometrical corrections. Such a goal was achieved, as we will argue next. 

In Section \ref{sec:EC} we have applied the energy conditions to the material content of our WH solutions. From Figures \ref{fig1}-\ref{fig11}, we can see that, apart from DEC in the transverse pressure case, the energy conditions are satisfied for a wide range of values for the parameter $\beta$.

Our results are consistent with those of References \cite{zubair/2016,Yousaf/2017}. The authors in \cite{zubair/2016} studied a specific $f(R,T)$ model with $f(R)=R+\alpha R^2+\gamma R^n$, with $\gamma$ and $n$ constants. In \cite{Yousaf/2017}, the authors considered $f(R)=R+\alpha R^2$. For the same shape function, these references show the validity of NEC and WEC for a specific small range of $r$. In our trace of the energy-momentum squared gravity model, both NEC and WEC are satisfied for a wide range of values for $r$ and free parameters. This advantage of our results is due to the squared trace of the energy-momentum tensor dependence of the theory. The possibility of such an attainment in EGTs that contain both geometrical and material correction terms shall be reported soon in the literature.

\acknowledgments PHRSM would like to thank S\~ao Paulo Research Foundation (FAPESP), grant 2015/08476-0, for financial support. The authors thank the referee for his/her valuable report. It has contributed for the enrichment of the physical and mathematical content of the article, as well as for their presentation form. The authors are also thankful to M. Chinaglia and D.A.T. Vanzella for some enlightening discussions regarding the features of a space-time where $T_{\mu\nu}=diag(\rho,-p_r,-p_t,-p_t)$ holds.

\end{document}